\documentclass{appolb}
\usepackage{epsfig}
\newcommand \beq{\begin{eqnarray}}
\newcommand \eeq{\end{eqnarray}}

\begin{document}
\title{Weakly and strongly coupled degrees of freedom in the quark-gluon plasma%
\thanks{Talk presented at the workshop ``Excited QCD 2011", Les Houches, France,  Feb. 20-25, 2011}%
}
\author{Jean-Paul Blaizot
\address{IPhT, CNRS/URA 2306, CEA Saclay, 91191 Gif-sur-Yvette cedex, France}
}
\maketitle
\begin{abstract}
In this talk, I reflect on the physical origin of the strongly coupled character of the quark-gluon plasma.
\end{abstract}
  
\section{Introduction}

The data from RHIC reveals that the matter produced in ultra-relativistic heavy ion collisions is strongly interacting: this is manifest in the strong absorption of jets, commonly attributed to a large energy loss in matter, and  the collective behavior responsible for the elliptic flow \cite{RHIC}. Particularly striking, in the latter case,  is the success of hydrodynamics, suggesting perfect fluid behavior with a low ratio of viscosity to entropy density \cite{Luzum:2008cw}. Such features, which have been beautifully confirmed by the recent experiments at the LHC,  appear incompatible with the properties of the quark-gluon plasma that one may naively deduce from  QCD asymptotic freedom.

I would like to argue in this talk that the origin  of  this  strongly coupled character of the quark-gluon plasma remains in fact somewhat of  a puzzle. Is the initial concept wrong?  Is the coupling constant large? The answer to these  questions is negative, and actually the puzzle cannot be solved by referring solely to the strength of the interaction. Thus in particular, in spite of the valuable insight that they provide on the dynamics of strongly coupled gauge theories, the strong coupling expansion methods based on the AdS/CFT correspondence \cite{Gubser:2009fc} do not help clarifying the issue.  In fact, both the strongly coupled quark-gluon plasma picture provided by this correspondence, as well as the naive picture of a non interacting plasma, are extreme idealizations that both miss an essential property of the physical plasmas: that modes with different wavelengths are differently coupled. In a physical plasma, there are modes with long wavelengths which are strongly coupled, however weak the coupling may be, while short wavelength modes, which dominate the thermodynamics, may be weakly coupled if the coupling constant is not too large.

\section{Is initial concept wrong ? }

The initial concept for the (weakly) interacting quark-gluon plasma emerges naturally  from QCD asymptotic freedom, and an elementary renormalization group argument: When calculating thermodynamical functions, the natural scale $Q$ that enters  the running coupling  $\alpha_s\approx 1/\ln(Q/\Lambda_{QCD})$ is $Q\simeq 2\pi T$, where $T$ is the temperature. Thus, interactions are small when 
 when $T\gg \Lambda_{QCD}$ and matter becomes  ÇÊsimpleÊÈ, turning asymptotically into an ideal gas of quarks and gluons. It is important to realize however that this statement holds only for the degrees of freedom that dominate the thermodynamics, namely quarks and gluons with momenta of order $T$.  It does not apply to long wavelength modes, as we shall discuss later. 

So, is this initial concept wrong? No, QCD asymptotic freedom works, as evidenced by numerous calculations. 

In particular,  first principle QCD calculations, namely lattice calculations of the thermodynamical properties, provide good evidence that the transition from the hadronic world to the quark gluon plasma is not a phase transition proper, but a smooth crossover \cite{Aoki:2006we}. The various thermodynamic functions (pressure, entropy or energy density) are thus analytic, but at the transition, they exhibit a rapid variation with  temperature (corresponding essentially to the ``liberation'' of quark and gluon degrees of freedom) and go, as expected, towards their free particle limit when the temperature becomes very large.  New  techniques have allowed recently  lattice calculations to be performed at arbitrarily high temperature, and demonstrate indeed the approach to the Stefan-Boltzmann  limit in a convincing way, in good agreement with weak coupling calculations \cite{Endrodi:2007tq}.

The calculation of the fluctuations of conserved charges (B,Q,S) provides evidence that, soon beyond the deconfinement transition, the response to small changes in the corresponding chemical potentials is quantitatively comparable to the response of weakly interacting quarks \cite{Cheng:2008zh}.

There are also attempts to reconstruct the spectral functions of quark excitations of the plasma. Although such calculations are difficult and not without ambiguities, they reveal a quasiparticle structure that is very much reminiscent of what weak coupling techniques lead us to expect  \cite{Karsch:2009tp}.

Finally, appropriate resummations of  QCD perturbation theory (see e.g. \cite{Blaizot:2000fc,Andersen:2010wu})  reproduce lattice results for temperatures greater than 2.5 to 3 Tc. This also  strongly suggests that the dominant effect  of the interactions in the quark-gluon plasma is to turn the massless quarks and gluons into weakly interacting massive quasiparticles.

One may argue that,  at RHIC, the quark-gluon plasma spends most of its existence in a region, say between  $T_c$ and  $\sim  3T_c$, where the physics is hard and poorly understood. It seems indeed that, in this region, the quasiparticle picture breaks down, and genuine non-perturbative effects appear in bulk thermodynamics.  Still, even in that region, explicit calculations reveal that the coupling constant is not huge, as we now discuss.

\section{Is the coupling constant large ?}

Not really. The running of the QCD coupling is well determined, both theoretically, and experimentally, so we know its value in the range of scales of interest. At the GeV scale it is not small, but it is not huge either,  $\alpha_s\approx .3 $ to $.4$. These are values that are indeed not small enough to guarantee the high accuracy that prevails for instance in high order perturbative QCD calculations  at the Z-boson scale. However, precision is not the issue here. We know that with a coupling constant $\alpha_s\approx .3 $ one is not misled badly when using perturbation theory to calculate a variety of low energy hadronic observables. 

In high temperature calculations, the relevant coupling constant can be estimated, and it is found to be of the  order of magnitude expected at the GeV scale \cite{Laine:2005ai}.
Of course, because $g^2=\alpha_s/4\pi$, a value of $\alpha_s\approx .3$ yields $g\approx 2$, which is not small compared to unity. This ``large'' value of the gauge coupling $g$ is often used as an argument against the usefulness of weak coupling techniques at finite temperature. We shall see however that the situation there is more subtle. 

\section{About the breakdown of strict perturbation theory}

 Much effort has been put into calculating
 the successive orders of the perturbative expansion for the pressure
and
the series is known now up to order $g^6\ln
g$ (see \cite{Kajantie:2002wa} and references therein).  These calculations have revealed  that strict 
perturbation theory  makes sense only for very small values of the
coupling constant, corresponding to extremely large values of $T$.
For not too small values of the coupling, the successive terms in
the expansion oscillate wildly and the dependence of the results on
the renormalization scale keeps increasing order after order (see e.g.
\cite{Blaizot:2003tw}), making strict perturbation theory inapplicable to estimate the corrections to the ideal quark
gluon plasma. 

 This situation is to be contrasted with what happens at zero temperature
 where, as we just said,  perturbative calculations provide reasonable guidance already at the GeV scale.
 The point is that  the validity of the weak coupling expansion depends
not only on the strength of the coupling, but also on the number of
active degrees of freedom. At zero temperature, one deals most of
the time with a very limited number of degrees of freedom (the
colliding particles and the reaction products), while at finite
temperatures, the  thermal fluctuations
alter the infrared behavior in a profound way. 
Let us emphasize that this problem is to a large extent not specific to QCD, but also occurs in simpler scalar field theories (see  \cite{Blaizot:2003tw}).

In the quark-gluon plasma, the effect of the interactions at a given scale $\kappa$ depends
on the magnitude of the  thermal fluctuations of the gauge fields  at that scale, $\langle A^2\rangle_\kappa$.  A simple analysis leads us to expect the expansion parameter (ratio of potential to kinetic energy)  to be of the form 
\beq
\gamma_\kappa=\frac{g^2 \langle A^2\rangle_\kappa}{\kappa^2}\sim\frac{g^2T}{\kappa}.
\eeq
 The fluctuations that dominate the energy density at weak coupling correspond to the plasma particles and have momenta $k\sim T$. For these ``hard'' fluctuations, 
$
 \gamma_T\sim g^2$, so that, at this scale,  perturbation
theory works as well as at zero temperature (with expansion parameter $\sim g^2$, or rather $\alpha=g^2/4\pi$). 
The next natural scale, commonly referred to as the ``soft scale'', corresponds to $\kappa\sim gT$. Then $ \gamma_{gT}\sim g$, so that perturbation theory can still be used to describe the self-interactions of the soft modes, however  it is now an expansion in powers of $g$ rather than $g^2$: it is therefore less rapidly convergent. 
Another phenomenon occurs at the scale $gT$. While the expansion parameter $ \gamma_{gT}$ that controls the self-interactions of the soft fluctuations is small, the coupling between the soft modes and thermal fluctuations at scale $T$ is not: indeed $g^2\langle A^2\rangle_T\sim (gT)^2$. Thus the dynamics of soft modes is non-perturbatively renormalized by their coupling to hard modes. This particular coupling is encompassed by the so-called hard thermal loops \cite{HTL}. 
Finally, there is yet another scale, the ``ultra-soft scale'' $\kappa\sim  g^2 T$, at which perturbation theory completely breaks down. At this scale, we have indeed
$ \gamma_{g^2T}\sim 1$.
Thus the ultra-soft  fluctuations remain strongly coupled for arbitrarily small couplings. This is what happens for the long wavelength, unscreened, magnetic
fluctuations. 

These considerations suggest that  the  main difficulty with thermal perturbation
theory is 
not so much related to the fact that the coupling is not small
enough (for the relevant temperatures it is not huge, as we have already pointed out), but
rather to the interplay of degrees of freedom with various
wavelengths, possibly involving collective modes.  One may object that the analysis above requires $g$ to be a number smaller than 1 to be meaningful. But the (exact) renormalization group allows one to easily overcome this apparent limitation of the strict weak coupling analysis  (see for instance \cite{Blaizot:2010ut} for an explicit calculation in scalar field theory).

\section{Initial stages of heavy ion collisions}

Finally, I would like to emphasize that  our present understanding of the initial stages of ultra-relativistic heavy ion collisions relies heavily on weak coupling considerations. Let us recall that 
the wave function of a relativistic system  describes a collection of partons, mostly gluons, whose number grows  when the system is boosted to higher energy  (then $x$, the typical momentum fraction carried by a gluon, decreases). 
 One
expects, however, that the growth of the gluon density 
eventually ``saturates'' when non linear QCD effects start to play a
role. The existence of such a saturation regime has been predicted long
ago, but it is only during the last
decade that equations providing a dynamical description of this regime have been obtained (for recent reviews, see \cite{Iancu:2003xm}).

The onset of saturation is characterized by a particular 
momentum scale, called the saturation momentum  $Q_s$.
Partons in the wave function have different transverse momenta
$k_T$. Those with  $k_T> Q_s$ are in a dilute regime; those with
$k_T<Q_s$ are in the saturated regime. Note that at saturation,
naive perturbation theory breaks down, even though $\alpha_s(Q_s)$
may be small if $Q_s$ is large: the saturation regime is a regime of
weak coupling, but the large gluon density induces non perturbative effects (in a way somewhat reminiscent to what happens at finite temperature). 
The color glass
formalism is an effective theory that  provides a complete description of the evolution of the wave function as a function of energy \cite{Iancu:2003xm}. 

The saturation momentum increases as the gluon density increases.
This increase of the gluon density  may come from the decrease of $x$ with increasing energy ($Q_s^2\sim x^{-0.3}$), or  from the
additive contributions of several nucleons in a nucleus, so that 
$Q_s^2\propto\alpha_s A^{1/3}$, where $A$ is the number of nucleons
in the nucleus. Thus, the saturation regime sets in earlier (i.e.,
at lower energy) in collisions involving large nuclei than in those
involving protons. In a nucleus-nucleus collision, most partons that play a direct role in particle production have momenta of the order of $Q_s$. A very successful phenomenology based on the saturation picture has been developped at RHIC (see e.g. \cite{Gelis:2010nm} for recent reviews). 

It is true that the picture is still incomplete. In particular, understanding how the  quark-gluon plasma is produced, i.e., understanding  the detailed  mechanisms by which the partonic degrees of freedom of the wavefunctions  get freed and subsequently interact to lead to a thermalized system, remains a challenging problem. But this is a problem that, presumably,  should  find its solution within this framework.

\section{Summary}

 The strongly coupled character of the quark-gluon plasma does not seem related in any obvious way to a large value of the coupling constant.
Non perturbative features may arise from the cooperation of many degrees of freedom, or strong classical fields, with examples provided by the high temperature plasma and the color glass condensate.

\end{document}